# A new well behaved class of charge analogue of Adler's relativistic exact solution

## Mohammad Hassan Murad

**Abstract** The paper presents a new class of parametric interior solutions of Einstein–Maxwell field equations in general relativity for a static spherically symmetric distribution of a charged perfect fluid with a particular form of electric field intensity. This solution gives us wide range of parameter, $K$ ($0.69 \leq K \leq 7.1$), for which the solution is well behaved hence, suitable for modeling of superdense star. For this solution the gravitational mass of a superdense object is maximized with all degree of suitability by assuming the surface density of the star equal to the normal nuclear density $\rho_{\text{nm}} = 2.5 \times 10^{17} \text{kg m}^{-3}$. By this model we obtain the mass of the Crab pulsar $M_{\text{Crab}} = 1.401 M_\odot$ and the radius, $R_{\text{Crab}} = 12.98$ km constraining the moment of inertia $I_{NS,45} > 1.61$ for the conservative estimate of Crab nebula mass $2M_\odot$ and $M_{\text{Crab}} = 2.0156 M_\odot$ with radius, $R_{\text{Crab}} = 14.07$ km constraining the moment of inertia $I_{NS,45} > 3.04$ for the newest estimate of Crab nebula mass $4.6M_\odot$ which are quite well in agreement with the possible values of mass and radius of Crab pulsar. Besides this, our model yields the moments of inertia for PSR J0737-3039A and PSR J0737-3039B are $I_A = 1.4624 \times 10^{45} \text{g cm}^2$ and $I_B = 1.2689 \times 10^{45} \text{g cm}^2$ respectively. It has been observed that under well behaved conditions this class of parametric solution gives us the maximum gravitational mass of causal superdense object $2.8020 M_\odot$ with radius 14.49 km, surface redshift 0.4319, charge $Q = 4.67 \times 10^{20} C$, and central density $\rho_c = 2.68 \rho_{\text{nm}}$.

M. H. Murad (✉)
Department of Natural Sciences,
Daffodil International University
102, Sukrabad, Mirpur Road, Dhaka-1207, Bangladesh.
E-mail: murad@daffodilvarsity.edu.bd

## 1. Introduction

Ever since the formulation of Einstein's gravitational field equation numbers of exact solutions have been found (Delgaty 1998, Stephani *et al.* 2003). Exact solutions of Einstein-Maxwell gravitational field equations are of vital importance in relativistic astrophysics.

Bonnor (Bonnor 1965) showed that the charge density can play an important part in equilibrium of large bodies which may possibly be able to halt gravitational collapse. Astrophysicists have been finding stable equilibrium solution for charged fluid spheres to construct the models of various astrophysical objects of immense gravity by considering the distinct nature of matter or radiation (energy-momentum tensor) present in them. Such models successfully explain the characteristics of massive objects like quasar, neutron star, pulsar, quark star, black-hole or other These stars are specified in terms of their masses as white dwarfs ($< 1.44 M_\odot$, Chandrasekhar limit), Strange quark star (possible maximum mass, $2M_\odot$), and Neutron star ($1.4 M_\odot - 2.9 M_\odot$). Of course neutron stars or any stars are not composed of perfect fluid. But such solutions may be used to make a suitable model of superdense object with charge matter.

Eventually, these exact solutions have received considerable attention due to some of the following reasons:

- ✓ A spherical body can remain in equilibrium under its own gravitation and electric repulsion, no internal pressure is necessary, if the matter present in the sphere carries certain modest electric charge density. The problem of the stability of a homogeneous distribution of matter containing a net surface charge was considered by Stettner (Cited in Whitman 1981). He showed that a fluid sphere of uniform density with a modest surface charge is more stable than the same system without charge. His solution is also stable towards an increase in the net surface charge. The electric charge weakens gravity to the extent of turning it into a repulsive field, as happens in the vicinity of a Reissener–Nordström singularity. Thus the gravitational collapse of a charged fluid sphere to a point singularity may be avoided (de Felice 1995).

- ✓ Charged solutions of Einstein-Maxwell equations are useful in the study of cosmic matter.
- ✓ Charged-dust (CD) models and electromagnetic mass models are expected to provide some clue about structure of an electron (Bijalwan 2011).
- ✓ Several fluid spheres which do not satisfy some or all the relevant physical conditions i.e. reality conditions, become relevant when they are charged.

In this paper we have obtained variety of new parametric class of well behaved exact solutions of Einstein–Maxwell field equations by considering the metric form $g_{00} = e^\nu = B(1 + Cr^2)^2$ (Adler 1974, Adams 1975a, Kuchowicz 1975). Several nonsingular charged analogues of Adler's solution was obtained by Nduka (1976), Whitman and Burch (Whitman 1981), Pant and Rajasekhara (Pant 2011b).

To test the compatibility and well behaved nature of our solution we construct fluid spheres of various mass by setting different values of parameters. We apply this solution to find the maximum mass of superdence object, like neutron star, and calculate various physical parameters for the rotating compact objects like Crab Pulsar (PSR B0531-21, spin period $P$ = 33 ms), and recently discovered the first double pulsar system PSR J0737-3039 (Burgay 2003, Lyne 2004), composed of two active radio pulsars PSR J0737-3039A ($P$ = 22.69 ms) and PSR J0737-3039B ($P$ = 2.77 s) having precise gravitational mass (1.3381 ± 0.0007) $M_\odot$ and (1.2489 ± 0.0007) $M_\odot$ respectively (Kramer 2006a, 2006b). We show that our solution yields values that are quite well with the observations made in several recent researches.

## 2. Physical conditions for a regular and well behaved charged fluid sphere to construct a superdense star model

A spherically symmetric metric in curvature coordinates can be written as
$$ds^2 = e^\nu dt^2 - e^\lambda dr^2 - r^2(d\theta^2 + \sin^2\theta \, d\phi^2) \quad (2.1)$$

The functions $\nu(r)$ and $\lambda(r)$ satisfy the Einstein-Maxwell field equation
$$G^i_{\ j} = R^i_{\ j} - \frac{1}{2}R\delta^i_{\ j} = \kappa(T^i_{\ j} + E^i_{\ j}) \quad (2.2)$$

where we have chosen the units so that, $c = G = 1$, and $\kappa = 8\pi$ is *Einstein's constant*. $T^i_{\ j}$ and $E^i_{\ j}$ are the energy-momentum tensor of perfect fluid and electromagnetic field defined by,
$$T^i_{\ j} = (\rho + p)v^i v_j - p\delta^i_{\ j}$$
$$E^i_{\ j} = \frac{1}{4\pi}\left(-F^{im}F_{jm} + \frac{1}{4}\delta^i_{\ j}F^{mn}F_{mn}\right)$$

and $\rho, p, v^i, F_{ij}$ denote energy density, fluid pressure, velocity vector and anti-symmetric electromagnetic field strength tensor respectively. On account of the high nonlinearity of Einstein–Maxwell field equations not many realistic well behaved analytic solutions are known for the description of relativistic fluid spheres. For a well behaved model of a relativistic star with charged perfect fluid matter, following physical conditions should be satisfied (Sabbadini 1973, Glass 1978, Hartle 1978, Buchdahl 1979, Delgaty 1998):

(i) The solution should be free from physical and geometric singularities i.e. $e^\nu > 0$ and $e^\lambda > 0$ in the range $0 \leq r \leq R$

(ii) The hydrostatic pressure $p$, at zero temperature is a function of $\rho$ only, i.e. $p = p(\rho)$.

(iii) The pressure and density are positive, $p, \rho \geq 0$, where the last inequality is the statement that gravity is attractive.

(iv) Pressure $p$ should be zero at boundary $r = R$.

(v) In order to have an equilibrium star the matter must be stable against the collapse of local regions. This requires, *Le Chatelier's principle* also known as *local* or *microscopic stability* condition, that $p$ must be a monotonically non-decreasing function of $\rho$ (Rhoades 1974, Hegyi 1975),
$$\frac{dp}{d\rho} \geq 0$$

(vi) The quantity $\sqrt{\frac{dp}{d\rho}}$ is the hydrodynamic phase velocity of sound waves in the neutron star matter. In the absence of dispersion and absorption it would be the velocity of signals in the medium. Then the condition $\sqrt{\frac{dp}{d\rho}} \leq 1$ would then be the condition that the speed of these signals not exceeds that of light (*causality condition*).

(vii) $\rho \geq p > 0$ or $\rho \geq 3p > 0$, $0 \leq r < R$, where former inequality denotes weak energy condition (WEC), while the later inequality implies strong energy condition (SEC).

(viii) $\left(\frac{dp}{dr}\right)_{r=0}, \left(\frac{d\rho}{dr}\right)_{r=0} = 0$ and $\left(\frac{d^2p}{dr^2}\right)_{r=0}, \left(\frac{d^2\rho}{dr^2}\right)_{r=0} < 0$ so that pressure and density gradients $\frac{dp}{dr}, \frac{d\rho}{dr} < 0$ for $0 < r \leq R$.

The above condition implies that pressure and density should maximum at the centre and monotonically decreasing towards the pressure free interface (boundary of the sphere).

(ix) The velocity of sound should be decreasing towards the surface i.e. $\left(\frac{d}{dr}\left(\frac{dp}{d\rho}\right)\right)_{r=0} < 0$ for $0 \leq r \leq R$ or the velocity of sound is increasing with the increase of density.

(x) The ratio of pressure to the density $\frac{p}{\rho}$ should be monotonically decreasing with the increase of $r$ i.e. $\left(\frac{d}{dr}\left(\frac{p}{\rho}\right)\right)_{r=0} < 0$ and $\left(\frac{d^2}{dr^2}\left(\frac{p}{\rho}\right)\right)_{r=0} < 0$ and $\frac{d}{dr}\left(\frac{p}{\rho}\right)$ is negative valued function for $r > 0$.

(xi) Gravitational redshift, $z$, should be monotonically decreasing toward the boundary of the sphere. The central red shift $z_0$ and surface red shift $z_R$ should be positive and finite i.e. $z_c = \sqrt{e^{-\nu(0)}} - 1 > 0$ and $z_R = \sqrt{e^{\lambda(R)}} - 1 > 0$ and both should be bounded.

(xii) Electric field intensity $E$, such that $E(0) = 0$, is taken to be monotonically increasing i.e. $\frac{dE}{dr} > 0$ for $0 < r < R$.

(xiii) The relativistic adiabatic index is given by $\gamma = \frac{(p+\rho)}{p}\frac{dp}{d\rho}$. The necessary condition for an exact solution to serve as a model of a relativistic star is that $\gamma > \frac{4}{3}$ (Ipser 1970, Adams 1975b, Knutsen 1988a, 1988b, 1989, 1991).

For a given radius, a static fluid sphere can not have an arbitrary mass. Buchdahl (1959) has obtained an absolute constraint of the maximally allowable mass–radius ($M/R$) ratio for isotropic fluid spheres of the form $\frac{2M}{R} \leq \frac{8}{9}$ (in natural units, $c = G = 1$). Böhmer and Harko (Böhmer 2007) proved that for a general relativistic compact charged object with charge $Q$ there is a lower bound for the mass–radius ratio,

$$\frac{3}{2}\frac{Q^2}{R^2}\frac{\left(1 + \frac{Q^2}{18R^2}\right)}{\left(1 + \frac{Q^2}{12R^2}\right)} \leq \frac{2M}{R}$$

Andréasson (2009) generalized Buchdahl inequality for the charged case and proved that

$$\sqrt{M} \leq \frac{\sqrt{R}}{3} + \sqrt{\frac{R}{9} + \frac{Q^2}{3R}}$$

We combine these results to constrain the mass $M$ for a charged object with given radius $R$, and charge $Q$ ($< M$),

$$\frac{Q^2}{R^2}\frac{(18R^2 + Q^2)}{(12R^2 + Q^2)} \leq \frac{2M}{R}$$
$$\leq 2\left(\frac{R + \sqrt{R^2 + 3Q^2}}{3R}\right)^2 \quad (2.1)$$

In the forthcoming sections we shall use the following theorem (Pant 2011c) for showing the monotonically decreasing or increasing nature of various physical parameters for well behaved nature of our solution.

**Theorem 2.1** If $f(r) = g(x)$; $\left(\frac{dg}{dx}\right)_{x=0}$ and $\left(\frac{d^2g}{dx^2}\right)_{x=0}$ are non zero finite where $x = Cr^2, C > 0$, then,

Maxima of $f(r)$ exists at $r = 0$ if $\left(\frac{dg}{dx}\right)_{x=0}$ is finitely negative

Minima of $f(r)$ exists at $r = 0$ if $\left(\frac{dg}{dx}\right)_{x=0}$ is finitely positive

## 3. Einstein–Maxwell equation for charged fluid distribution

In view of the metric (2.1), the field equation (2.2)

gives (Nduka 1976, Dionysiou 1982),

$$\frac{v'}{r}e^{-\lambda} - \frac{(1-e^{-\lambda})}{r^2} = \kappa p - \frac{q^2}{r^4} \quad (3.1)$$

$$\left(\frac{v''}{2} - \frac{v'\lambda'}{4} + \frac{v'^2}{4} + \frac{v'-\lambda'}{2r}\right)e^{-\lambda} = \kappa p + \frac{q^2}{r^4} \quad (3.2)$$

$$\frac{\lambda'}{r}e^{-\lambda} + \frac{(1-e^{-\lambda})}{r^2} = \kappa \rho + \frac{q^2}{r^4} \quad (3.3)$$

where, prime (') denotes the differentiation with respect to $r$ and $q(r)$ represents the total charge contained within the sphere of radius $r$ defined by

$$q(r) = 4\pi \int_0^r \rho e^{\frac{\lambda}{2}} r^2 \, dr$$

Now let us assume

$$\left.\begin{array}{l} e^v = B(1+Cr^2)^2 \\ e^{-\lambda} = Z, \ x = Cr^2 \end{array}\right\} \quad (3.4)$$

Putting these transformations into (3.1) and (3.3), the equations become,

$$\frac{\kappa}{C}p = \frac{1+5x}{x(1+x)}Z - \frac{1}{x} + \frac{Cq^2}{x^2} \quad (3.5)$$

$$\frac{\kappa}{C}\rho = -2\frac{dZ}{dx} - \frac{Z}{x} - \frac{1}{x}\left(\frac{Cq^2}{x} - 1\right) \quad (3.6)$$

and (3.2) becomes,

$$\frac{dZ}{dx} - \frac{1+x}{x(1+3x)}Z = \frac{(1+x)}{x(1+3x)}\left(\frac{2Cq^2}{x} - 1\right) \quad (3.7)$$

The solution of the above differential equation is

$$Z\frac{(1+3x)^{\frac{2}{3}}}{x}$$

$$= \int \left\{\frac{(1+x)}{x^2(1+3x)^{\frac{1}{3}}}\left(\frac{2Cq^2}{x} - 1\right)\right\} dx + A \quad (3.8)$$

where $A$ is an arbitrary constant of integration.

To integrate (3.8) we assume,

$$\frac{E^2}{C} = \frac{Cq^2}{x^2} = \frac{K}{2}x(1+x)^2(1+3x)^{\frac{1}{3}} \quad (3.9)$$

where $K \geq 0$. The electric field intensity is so assumed that the model is physically significant and well behaved. In view of (3.9), (3.8) yields the following solution,

$$Z = 1 + \frac{K}{4}\frac{x(1+x)^4}{(1+3x)^{\frac{2}{3}}} + A\frac{x}{(1+3x)^{\frac{2}{3}}} \quad (3.10)$$

Using (3.9) and (3.10) into (3.5) and (3.6), the pressure and energy density expressions become,

$$\frac{\kappa}{C}p = \frac{K}{4}(1+x)^2\left[\frac{1+8x+11x^2}{(1+3x)^{\frac{2}{3}}}\right] + \frac{4}{(1+x)}$$

$$+A\frac{(1+5x)}{(1+x)(1+3x)^{\frac{2}{3}}} \quad (3.11)$$

and

$$\frac{\kappa}{C}\rho = -\frac{K}{4}(1+x)^2\left[\frac{3+21x+57x^2+47x^3}{(1+3x)^{\frac{5}{3}}}\right]$$

$$-A\frac{(3+5x)}{(1+3x)^{\frac{5}{3}}} \quad (3.12)$$

## 4. Properties of new class of solution

The central values of pressure and density are given by,

$$\frac{\kappa}{C}p_c = \frac{K}{4} + 4 + A$$

$$\frac{\kappa}{C}\rho_c = -\frac{3K}{4} - 3A$$

For $p_c$ and $\rho_c$ must be positive and $\frac{p_c}{\rho_c} \leq 1$ we have,

$$-\frac{K}{4} - 4 \leq A \leq -\frac{K}{4} - 1 \quad (4.1)$$

Differentiating (3.11) and (3.12) with respect to $x$, we obtain the pressure and density gradients,

$$\frac{\kappa}{C}\frac{dp}{dx} = \frac{K}{2}\frac{(1+x)(4+29x+72x^2+55x^3)}{(1+3x)^{\frac{5}{3}}} - \frac{4}{(1+x)^2}$$

$$+2A\frac{(1-5x^2)}{(1+x)^2(1+3x)^{\frac{5}{3}}} \quad (4.2)$$

$$\frac{\kappa}{C}\frac{d\rho}{dx}$$

$$= -\frac{K}{2}\frac{(1+x)(6+69x+255x^2+411x^3+235x^4)}{(1+3x)^{\frac{8}{3}}}$$

$$+10A\frac{(1+x)}{(1+3x)^{\frac{8}{3}}} \quad (4.3)$$

The causality condition is given by taking the square root after dividing (4.2) by (4.3) and for the values of $K \geq 0$ and $A$ satisfied by (4.1) the following must be satisfied

$$0 \leq \left(\frac{dp}{d\rho}\right)_{x=0} \leq 1$$

$$\left.\frac{d}{dx}\left(\frac{p}{\rho}\right)\right|_{x=0} < 0$$

and

$$\left.\frac{d}{dx}\left(\frac{E^2}{C}\right)\right|_{x=0} = \frac{K}{2} \geq 0$$

$$\left.\frac{d^2}{dx^2}\left(\frac{E^2}{C}\right)\right|_{x=0} = 3K \neq 0$$

By the theorem 2.1, mentioned in section 2, the above two inequalities show that electric field intensity $E$ is minimum at the centre and monotonically increasing.

## 5. Physical Boundary Conditions

Besides the above, the charged fluid spheres so obtained are to be matched over the boundary with Reissner-Nordström metric (Dionysiou 1982):

$$ds^2 = \left(1 - \frac{2m(r)}{r} + \frac{q^2}{r^2}\right)dt^2$$
$$- \left(1 - \frac{2m(r)}{r} + \frac{q^2}{r^2}\right)^{-1} dr^2$$
$$- r^2(d\theta^2 + \sin^2\theta \, d\phi^2) \quad (5.1)$$

which requires the continuity of , $e^\nu$ $e^\lambda$ and $q$ across the boundary $r = R$.

At the boundary

$$e^{\nu(R)} = \left(1 - \frac{2M}{R} + \frac{Q^2}{R^2}\right) \quad (5.2)$$

$$e^{-\lambda(R)} = \left(1 - \frac{2M}{R} + \frac{Q^2}{R^2}\right) \quad (5.3)$$

$$m(R) = M$$
$$q(R) = Q$$
$$p(R) = 0$$

where $M$, $R$ and $Q$ represent the total mass, radius and the total charge inside the fluid sphere respectively.

Now using $r = R, x = CR^2 = X$ and $p(R) = 0$ into (3.11) we can compute the arbitrary constant $A$

$$A = -\frac{K}{4}(1+X)^3 \frac{(1 + 8X + 11X^2)}{(1 + 5X)}$$
$$- \frac{4(1 + 3X)^{\frac{2}{3}}}{(1 + 5X)} \quad (5.4)$$

Using (5.3) and (3.10), (5.2) and (3.4) we can construct the following mass expression and the constant $B$ respectively, as

$$\frac{2M}{R} = \frac{KX(1+X)^2\{-1 + 5X^2\}}{4(1+3X)^{\frac{2}{3}}}$$
$$- A\frac{X}{(1+3X)^{\frac{2}{3}}} \quad (5.5)$$

$$B = \frac{K}{4}\frac{X(1+X)^2}{(1+3X)^{\frac{2}{3}}} + \frac{1}{(1+X)^2}$$
$$+ A\frac{X}{(1+X)^2(1+3X)^{\frac{2}{3}}} \quad (5.6)$$

The expression for the gravitational redshift is given by

$$z = \sqrt{e^{-\nu}} - 1 = \frac{1}{\sqrt{B}(1+x)} - 1$$

The central red shift is given by

$$z_c = \sqrt{e^{-\nu(0)}} - 1 = \frac{1}{\sqrt{B}} - 1$$

which must be nonnegative, i.e.

$$\frac{1}{\sqrt{B}} - 1 > 0 \Rightarrow 0 < \sqrt{B} < 1$$

and

$$\left(\frac{dz}{dx}\right)_{x=0} = -\frac{1}{\sqrt{B}} < 0$$
$$\left(\frac{d^2z}{dx^2}\right)_{x=0} = \frac{2}{\sqrt{B}} \neq 0$$

The above two inequalities indicate that the gravitational redshift is maximum at the center (theorem 2.1) and monotonically decreasing towards the pressure free interface.

The boundary redshift is given by

$$z_R = \sqrt{e^{\lambda(R)}} - 1 = \frac{1}{\sqrt{B}(1+X)} - 1$$

Denoting the boundary surface density $\rho(R) = \rho_s$, (3.12) gives,

$$\kappa R^2 \rho_s = -\frac{KX}{4}(1+X)^2 \left[\frac{3 + 21X + 57X^2 + 47X^3}{(1+3X)^{\frac{5}{3}}}\right]$$
$$- AX\frac{(3+5X)}{(1+3X)^{\frac{5}{3}}} = L \quad (5.7)$$

Now the radius of the charged fluid sphere becomes

$$R = \sqrt{\frac{L}{\kappa \rho_s}} \quad (5.8)$$

## 6. Calculations and Tables of numerical values

To construct well behaved model of superdense astrophysical object we shall be using the following numerical values

Nuclear matter density as surface density,
$\rho_{nm} = \rho_s = 1.857 \times 10^{-10}\,\text{m}^{-2} = 2.5 \times 10^{17}\,\text{kg m}^{-3}$,
$c = 1 = 2.997 \times 10^8\,\text{ms}^{-1}$,
$G = 1 = 6.674 \times 10^{-11}\,\text{Nm}^2\text{kg}^{-2}$,
$M_\odot = 1.486\,\text{km} = 2 \times 10^{30}\,\text{kg}$

**Table 1** The variation of various physical parameters e. g. pressure, central density, pressure-energy density ratio, causality condition, surface redshift, and central density. [We shall follow the notation $\rho_{0,17} = \frac{\rho_0}{10^{17}\,\text{kg m}^{-3}}$]

| $K$ | $X$ | $\frac{\kappa}{C}p_c$ | $\frac{\kappa}{C}c^2\rho_c$ | $\frac{1}{c^2}\left(\frac{p}{\rho}\right)_c$ | $\sqrt{\frac{1}{c^2}\left(\frac{dp}{d\rho}\right)_c}$ | $z_R$ | $\frac{M_G}{M_\odot}$ | $R$ (km) | $\rho_{0,17}$ |
|---|---|---|---|---|---|---|---|---|---|
| 1.88 | 0.106 | 0.5078 | 10.4764 | 0.0484 | 0.4901 | 0.1845 | 1.2492 | 12.31 | 3.92 |
| 1.25 | 0.116 | 0.6524 | 10.0426 | 0.0649 | 0.5181 | 0.1968 | 1.3380 | 12.69 | 3.87 |
| 1.22 | 0.178 | 0.7585 | 9.7242 | 0.0780 | 0.5216 | 0.2867 | 1.9701 | 13.71 | 4.92 |
| 0.66 | 0.3 | 1.0739 | 8.7780 | 0.1223 | 0.3063 | 0.4319 | **2.8020** | 14.49 | 6.71 |
| 0.75 | 0.19 | 0.9439 | 9.1682 | 0.1029 | 0.5456 | 0.2946 | 2.0156 | 14.07 | 4.70 |
| 1.57 | 0.167 | 0.6364 | 10.0907 | 0.0630 | 0.5048 | 0.2759 | 1.9008 | 13.45 | 4.98 |
| 4 | 0.061 | 0.1721 | 11.4835 | 0.0149 | 0.3948 | 0.1139 | 0.7082 | 10.36 | 3.48 |
| 1.41 | 0.112 | 0.6102 | 10.1694 | 0.0600 | 0.5109 | 0.1916 | 1.3002 | 12.55 | 3.86 |
| 0.8 | 0.124 | 0.7724 | 9.6827 | 0.0797 | 0.5388 | 0.2057 | 1.4011 | 12.98 | 3.81 |

**Table 2** The variation of various physical parameters e. g. pressure, surface density, pressure-energy density ratio, causality condition, gravitational redshift, pressure gradient, density gradient and relativistic adiabatic index in the fluid sphere with, $K = 0.66$, $X = 0.3$

| $\frac{r}{R}$ | $\frac{\kappa}{C}p$ | $\frac{\kappa}{C}c^2\rho$ | $\frac{1}{c^2}\frac{p}{\rho}$ | $\frac{\kappa}{C}\frac{dp}{dx}$ | $\frac{\kappa c^2}{C}\frac{d\rho}{dx}$ | $\sqrt{\frac{1}{c^2}\frac{dp}{d\rho}}$ | $Q$ (km) | $z$ | $\gamma$ |
|---|---|---|---|---|---|---|---|---|---|
| 0 | 1.073983 | 8.778052 | 0.122349 | -8.86203 | -32.8902 | 0.519079216 | 0 | 0.861568 | 4.761703 |
| 0.1 | 1.047645 | 8.680305 | 0.120692 | -8.69694 | -32.2776 | 0.519077811 | 0.025095 | 0.856 | 4.819919 |
| 0.2 | 0.971538 | 8.397693 | 0.115691 | -8.22014 | -30.5522 | 0.518702622 | 0.101727 | 0.839494 | 5.002217 |
| 0.3 | 0.853857 | 7.958965 | 0.107282 | -7.48178 | -28.0096 | 0.516831366 | 0.233931 | 0.812627 | 5.334314 |
| 0.4 | 0.706711 | 7.4031 | 0.095461 | -6.55055 | -25.0316 | 0.511557748 | 0.428407 | 0.776306 | 5.870347 |
| 0.5 | 0.544383 | 6.770139 | 0.080409 | -5.49815 | -21.9821 | 0.500119604 | 0.694505 | 0.731691 | 6.719782 |
| 0.6 | 0.381747 | 6.093907 | 0.062644 | -4.38686 | -19.1401 | 0.478745956 | 1.044226 | 0.680116 | 8.121071 |
| 0.7 | 0.233157 | 5.398079 | 0.043193 | -3.26269 | -16.6793 | 0.442281981 | 1.492249 | 0.622989 | 10.68205 |
| 0.8 | 0.111933 | 4.695265 | 0.02384 | -2.15366 | -14.6815 | 0.383004616 | 2.055985 | 0.561718 | 16.44897 |
| 0.9 | 0.03035 | 3.987977 | 0.00761 | -1.07162 | -13.1644 | 0.285312021 | 2.755664 | 0.497641 | 37.77477 |
| 1 | 0 | 3.270323 | 0 | -0.01552 | -12.1094 | 0.035804278 | 3.614431 | 0.431976 | ∞ |

## 7. An application of the model to the Crab pulsar, PSR J0737-3039, and PSR J1614-2230

The mass and the moment of inertia are the two gross structural parameters of neutron stars which are most accessible to observation. It is the mass which controls the gravitational interaction of the star with other systems such as a binary companion. It is the moment of inertia which controls the energy stored in rotation and thereby the energy available to the pulsar emission mechanism. Mass and moment of inertia have been using to constrain the dense matter equation of state (EOS) in the interior of neutron stars (Bejger 2005). For the Crab Pulsar (PSR B0531-21, spin period $P = 33$ ms) and the first double pulsar system, PSR J0737-3039, composed of two radio pulsars PSR J0737-3039A ($P = 22.69$ ms, $M_G = 1.338 M_\odot$) and PSR J0737-3039B ($P = 2.77$ s, $M_G = 1.249 M_\odot$), and one of the most recently discovered (Demorest 2010) massive binary millisecond pulsar PSR J1614-2230 ($P = 3.15$ ms, $M_G = 1.97 \pm 0.04 M_\odot$) we calculate the moment of inertia by the very precise, "empirical formula" which is based on the numerical results obtained for a thirty theoretical EOSs of dense nuclear matter (Bejger 2002),

$$I \simeq a(x) M R^2 \quad (7.1)$$

$$a(x) = \begin{cases} a_{NS}(x) = \begin{cases} \dfrac{x}{(0.1 + 2x)} & x \leq 0.1 \\ \dfrac{2}{9}(1 + 5x) & x > 0.1 \end{cases} \\ a_{SS}(x) = \dfrac{2}{5}(1 + x) \end{cases} \quad (7.2)$$

where $x$ is the dimensionless compactness parameter ($M$ and $M_\odot$ are measured in km).

$$x = \frac{\left(\dfrac{M}{R}\right)}{M_\odot} = \frac{\dfrac{M}{M_\odot}}{R}$$

Equations (7.1) and (7.2) are used to calculate the moment of inertia of several fluid spheres for the model considered in the present study.

**Table 3** Moments of inertia of various well behaved charged fluid spheres with known gravitational mass of pulsars [Notations, $I_{NS,45} = \dfrac{I_{NS}}{10^{45} \text{ kg m}^2}$, $Q_{20} = \dfrac{Q}{10^{20}C}$]

| K | X | $\dfrac{M_G}{M_\odot}$ | $I_{NS,45}$ | $Q_{20}$ | $\dfrac{Q}{M_G}$ | $\dfrac{\rho_c}{\rho_{nm}}$ |
|---|---|---|---|---|---|---|
| 1.88 | 0.106 | 1.2492 | 1.2689 | 2.08 | 0.7894 | 1.56 |
| 1.25 | 0.116 | 1.3380 | 1.4624 | 2.23 | 0.6864 | 1.54 |
| 1.22 | 0.178 | 1.9701 | 2.8301 | 3.28 | 0.8239 | 2.46 |
| 0.66 | 0.3 | 2.8020 | 5.1462 | 4.67 | 0.8680 | 2.68 |
| 0.75 | 0.19 | 2.0156 | 3.0455 | 3.35 | 0.7014 | 1.88 |
| 1.57 | 0.167 | 1.9008 | 2.6099 | 3.16 | 0.8801 | 1.99 |
| 1.41 | 0.112 | 1.3002 | 1.3827 | 2.16 | 0.7131 | 1.54 |
| 4 | 0.061 | 0.7082 | 0.4395 | 1.18 | 0.9272 | 1.39 |
| 0.8 | 0.124 | 1.4011 | 1.6156 | 2.33 | 0.5793 | 1.52 |

## 8. Discussion

In view of Table 1 we observe that all the physical parameters $\left(p, \rho, \dfrac{1}{c^2}\dfrac{p}{\rho}, \dfrac{1}{c^2}\dfrac{dp}{d\rho}, z\right)$ are positive at the centre and within the limit of realistic equation of state. In this article, Adler solution of Einstein's gravitational field equations in general relativity has been charged by means of suitable charge distribution $\dfrac{E^2}{C} = \dfrac{Cq^2}{x^2} = \dfrac{K}{2}x(1+x)^2(1+3x)^{\frac{1}{3}}$, as compared to that of Pant-Tewari solution (Pant 2011), and Pant-Tewari-Fuloria solution (Pant 2011a), which is zero at the center and monotonically increasing towards the pressure free interface. Our solution satisfies well behaved conditions for wide range of values of $K$ ($0.69 \leq K \leq 7.1$). The resulting fluid spheres can be utilized to construct the models of various compact astrophysical charged objects. Owing to the various conditions that we obtain here we arrive at the conclusion that, for the suitable choice of parameters, this class of solutions gives us the maximum gravitational mass of superdense object, $M_{G(\max)} = 2.8020 M_\odot$ with radius 14.49 km, surface redshift $z_R = 0.4319$, charge $Q = 4.67 \times 10^{20} C$, and central density $\rho_c = 6.71 \times 10^{17}$ kg m$^{-3}$ [$\rho_c = 2.68 \rho_{\text{nm}}$]. This maximum value, however, lies within the range of the upper limit of maximum mass of neutron star calculated by Kalogera and Baym (Kalogera 1996) employing WFF88 EOS with fiducial density $\rho_f = 4.6 \times 10^{17}$ kg m$^{-3}$.

Corresponding to the values $K = 0.8$, $X = 0.124$ (table 2, 3) we found a fluid sphere of gravitational mass $1.4011 M_\odot$ with radius and moment of inertia $R = 12.72$ km and $I_{\text{Crab},45} = 1.6101$. These values are quite well agreement with the possible mass and radius of the Crab pulsar, constraining the moment of inertia of the Crab pulsar $I_{\text{Crab},45} > 1.61$ for Crab Nebula mass $M_{\text{Neb}} = 2 M_\odot$ (conservative estimate). Moreover, corresponding to the values $K = 0.75$, $X = 0.19$ our model gives a fluid sphere of gravitational mass $2.0156 M_\odot$ with radius $R = 14.07$ km. These values are also quite well agreement with the predicted mass and radius of the Crab pulsar, constraining $I_{\text{Crab},45} > 3.04$ for $M_{\text{Neb}} = 4.6 M_\odot$ (newest estimate), in Bejger and Haensel (Bejger 2002).

Besides this, our model yields the moment of inertia $I_{A,45} = 1.4624$ ($K = 1.25$, $X = 0.116$) for PSR J0737-3039A and $I_{B,45} = 1.2689$ ($K = 1.88$, $X = 0.106$) for PSR J0737-3039B which are also within the range calculated by various EOSs in Bejger (2005), and Worley (2008). For the values $K = 1.41$, $X = 0.112$ we obtain $M_G = 1.3002 M_\odot$ with radius $R = 12.55$ km which are quite similar mass, $M = 1.3 \pm 0.6 M_\odot$, and radius, $R = 11^{+3}_{-2}$ of the neutron star in low-mass X-ray burster 4U 1820-30 (Kuśmierek 2011) but at some other report (Güver 2010) the mass and radius were measured, for the burster 4U 1820-30, $1.58 \pm 0.06 M_\odot$ and $9.1 \pm 0.4$ km. For $K = 2.12$, $X = 0.134$, our model gives the mass $M = 1.5884 M_\odot$ and radius 12.86 km.

In absence of the charge, ($K = 0$), however, we are left behind with the neutral Adler solution (component of exact metric, $g_{00}$, is same for $Q = 0$ and $Q > 0$), which is not well behaved as causality condition does not hold (Delgaty 1998, Kiess 2012).

**Acknowledgements** Author acknowledges his gratitude to Saba Fatema, Lecturer, Department of Natural Sciences, Daffodil International University, Dhaka, Bangladesh, for her help and continuous support. Author is also grateful to the reviewers and referees for pointing out the errors and making relevant constructive suggestions that helped author to improve the original manuscript.